\documentclass[journal]{IEEEtran}
\usepackage{blindtext}
\usepackage{graphicx}  
\usepackage{amsmath,amssymb}

\begin{document}
\title{Single-Multi-Single Fibre Optic Structure Based Water Depth Sensor }

\author{Garima~Bawa, Krishnendu~Dandapat, Gyanendra~Kumar, Indrajeet~Kumar and Saurabh~Mani~Tripathi, ~\IEEEmembership{Senior~Member,~IEEE}

\thanks{ G. Bawa, K. Dandapat and S. M. Tripathi are with the Physics Department, Indian Institute of Technology Kanpur, Kanpur 208016, India (e-mail: garimab@iitk.ac.in, krishnd@iitk.ac.in; tripathi.iit@gmail.com).
	
G. Kumar is with Indian Institute of Technology, Roorkee 247667, India (email: gyaniitk@gmail.com).

I. Kumar and S. M. Tripathi are with Center for Lasers and Photonics, Indian Institute of technology Kanpur, Knapur 2018016, India(e-mail: k.indrajeet@gmail.com; tripathi.iit@gmail.com)}
\thanks{Manuscript received xx, xxxx.}}

\markboth{}
{Kumar \MakeLowercase{\textit{et al.}}: Single-Multi-Single Fibre Optic Structure Based wavelength Division Multiplexing Channel Isolation Filter}
\maketitle

\begin{abstract}
We propose and demonstrate a technique easy to fabricate and measure water depth based on selective mode excitation using single-multi-single (SMS) mode structure. The high extinction ratio has been achieved by equally exciting preselected modes of the multi-mode fibre (MMF) by optimizing the core offset at both the input/output splices of the SMS structure, and a wavelength shifting has been achieved by varying the ratio of major to minor axis of the MMF loop. A theoretical analysis of the observed behaviour is also presented showing an excellent agreement between the theoretical and experimental results.
\end{abstract}


\IEEEpeerreviewmaketitle

\section{Introduction}
\IEEEPARstart{W}{avelength} division multiplexing (WDM) is widely used to handle numerous data streams to increase the transmission capacity of data communication network \cite{Okamoto}. In WDM, optical signals of various wavelengths are transmitted through a single optical fiber, owing to which it offers increased bandwidth without disturbing the globally embedded fiber network. At the output end of such network, WDM channel isolation filters with high loss between preselected adjacent channels (to reduce cross talks between these channels) are employed. Several schemes of WDM channel isolation filters have been reported in literature using chirped long period gratings \cite{Das, Tiwari}, Fibre Bragg gratings \cite{Dixit}, and uniform long period gratings \cite{Gu} etc. These filters, apart from offering their distinct advantages \cite{Tiwari, Dixit, Gu, Murthy}, also suffer from certain drawbacks including complex and costly fabrication techniques and the necessity of perfectly aligned splicing at both ends of the gratings to obtain well-shaped spectrum. To overcome these, often costly and complex design of fibre optic grating based WDM channel isolation filter, in this letter we present a new technique based on multi-modal interference (MMI) effect \cite{ArunK, Chen} in optical fibres by using single mode-multi mode-single mode (SMS) structure \cite{ArunK, SMT1}.

Modal intereference effect in SMS structures has been extensively used to develop several optical devices such as band pass/stop filters \cite{Mohammed, Lopez,Tripathi}, sensors \cite{Li, SMT1, SMT2}, modulators, directional couplers \cite{Mackie}, optical switches and power splitters etc. It has been an active area of research due to its low cost, ease of fabrication and possibility of easily modifying the structure (their length/type of the fibres etc.) as per the desired applications.

In a SMS structure, two identical single mode fibers (SMFs) are axially spliced with a small section of multi mode fibre (MMF) in between them. When both the splices are perfectly aligned, only first few symmetric modes of MMF get excited, with the fundamental mode carrying most of the power \cite{Murthy}. Owing to this, the extinction ratio of the transmission spectrum is quite low. To increase the extinction ratio we axially misaligned both of the SMF-MMF splices. As a result, a redistribution of power carried by individual modes of MMF takes place, with power coupled to the fundamental mode of MMF becoming close to that carried by the higher order modes, due to which the extinction ratio increases. To demonstrate the fast tuning properties of this WDM isolation filter we varied the ratio of major to minor axis of the MMF loop and recorded the transmission spectra. Our experiment has been carried out over the wavelength range of 1530-1570 nm, which is of prime importance in fibre optic communication system due to extremely low loss of the optical fibres over this wavelength range. 

\section{Experimental details}

\begin{figure}[ht!]
\begin{center}
\includegraphics [width=9 cm]{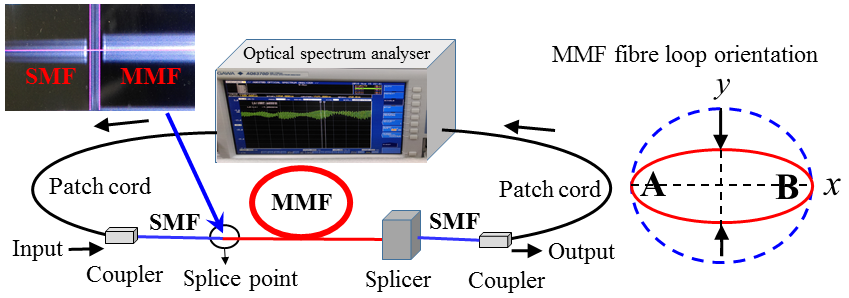}
\end{center}
\caption {Schematic diagram of the experimental set-up. SMF - Single mode fibre (SMF-28), MMF - Multi mode fibre (Corning). Insect shows the snapshot of misaligned lead-in splice point.}
\label{exptsetup}
\end{figure}

The schematic diagram of our experimental configuration is shown in Fig. \ref{exptsetup}. To built the filter, a long piece of MMF (Corning-F01GI62NCG01) having length of 6.5 $m$, coiled in a loop of diameter 8 $cm$, was spliced in between two SMFs (Sterlite-6CH5492). Both splice points were slightly misaligned to excite the higher order modes of MMF and the core offset was optimized to obtain high extinction ratio.The core and cladding diameters of the SMF and MMF used in our experiment are 8/125 $\mu m$ (Sterlite) and 50/125 $\mu m$ (Corning), respectively. The MMF was coiled to reduce the overall size of the filter. The used MMF has parabolic refractive index profile \cite{SMT1}. In our experiment, we adjusted the MMF length to get a channel isolation of 2.5 $nm$, which can be further increased/decreased by decreasing/increasing the MMF length \cite{Murthy}. One end of this MMF was misaligned in $x$,$y$-directions and spliced with lead-in SMF using fusion splicer (FUJIKURA 80S), as shown in Fig. \ref{exptsetup}. The other end of the MMF was misaligned in the $x$-direction and spliced with lead-out SMF to form SMS structure. Light was launched into the lead-in SMF through the inbuilt broadband source (LEUKOS, SM-30-450) and the transmission spectra were recorded using a high precision optical spectrum analyser (OSA, YOKOGAWA, AQ6370D).  Except the loop region of the MMF, we fixed the SMS structure to a platform such that the SMS structure remains straight to avoid any other shifts in the transmission spectrum due to strains of other regions of the SMS structure.

\section{Experimental results and discussion}

\begin{figure}[h!]
\begin{center}
\includegraphics [width=7 cm]{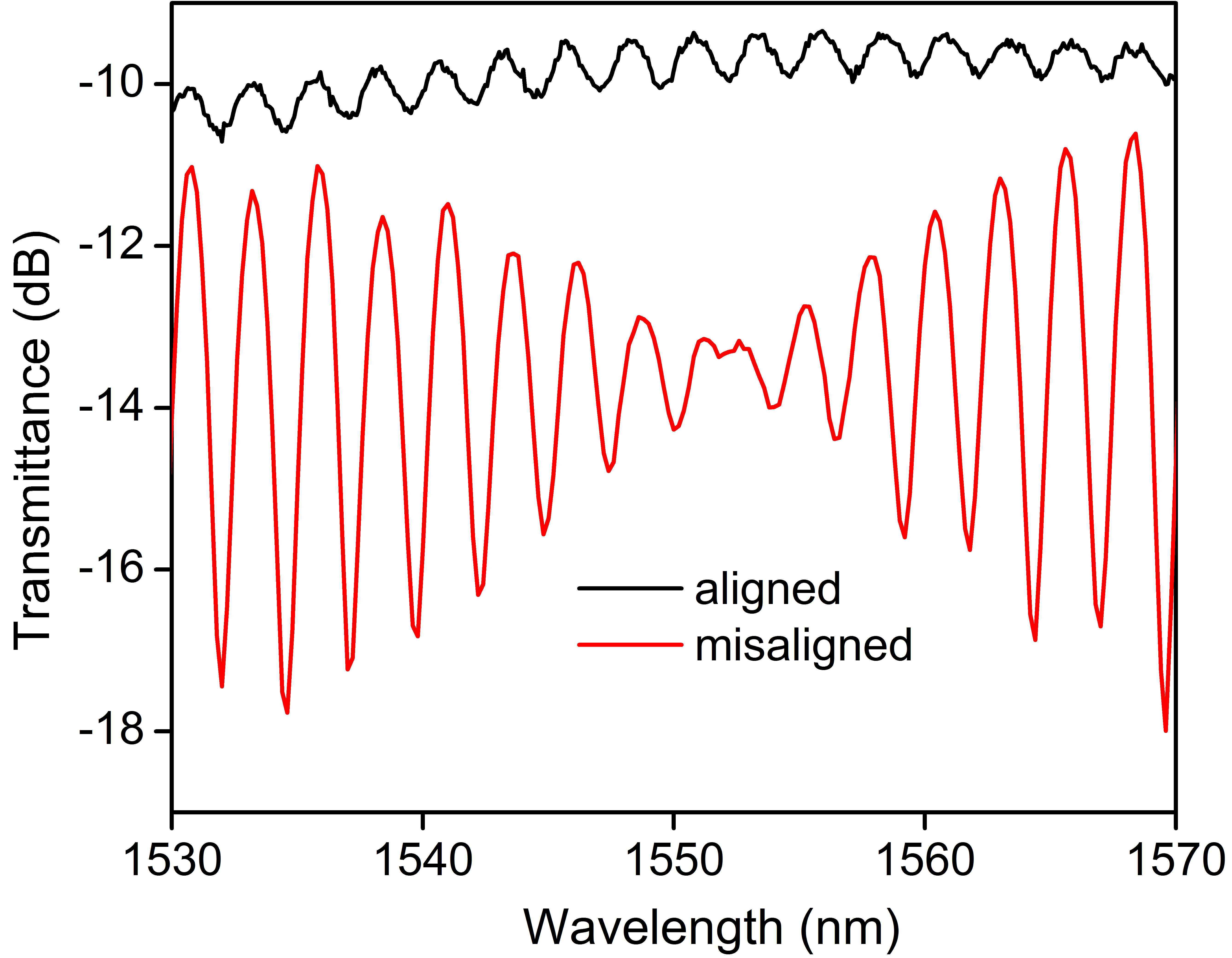}
\end{center}
\caption {Transmission spectrum of SMS structures when SMF-MMF are perfectly aligned (black line) and misaligned (red line) of MMF loop.}
\label{misalign}
\end{figure}

The transmission spectrum of the SMS structure for axially aligned and with axial misalignment (solid curve) have been plotted in Fig. \ref{misalign}. When the SMFs and MMF are perfectly aligned the extinction ratio is quite low, nearly 2 dB (shown by black line). In contrast to that when an axial misalignment of 2.7 $\mu m$ ((2, 5)-misalignment of 2 and 5 steps along $x$ and $y$-axis, respectively) at one splice point (lead-in) while 1.5 $\mu m$ ((3, 0)-misalignment of 3 steps along $x$-axis) at other splice point (lead-out) is introduced, the extinction ratio of the transmission spectrum is enhanced. This is due to nearly equal excitation efficiency of various MMF modes as discussed later. We fabricated several filters by first optimizing the offset at individual splices and later on by simultaneously optimizing the misalignments at both the splices. Fig. \ref{misalign} (solid curve) shows the transmission spectrum of optimized filter. Due to misaligned splices, the fraction power coupled to the $LP_{01}$ mode of the MMF decreased and becomes close to that carried by other modes of the MMF. As a result, modal interference occurs which leads to the enhancement in extinction ratio. A second order of optimization was further carried out by optimizing the major ($a$) to minor ($b$) axis ratio ($a/b$) to achieve highest extinction ratio. By changing the $a/b$, the extinction ratio was increased (shown by solid red line in Fig. \ref{misalign}).

\begin{figure}[h!]
\begin{center}
\includegraphics [width=9 cm]{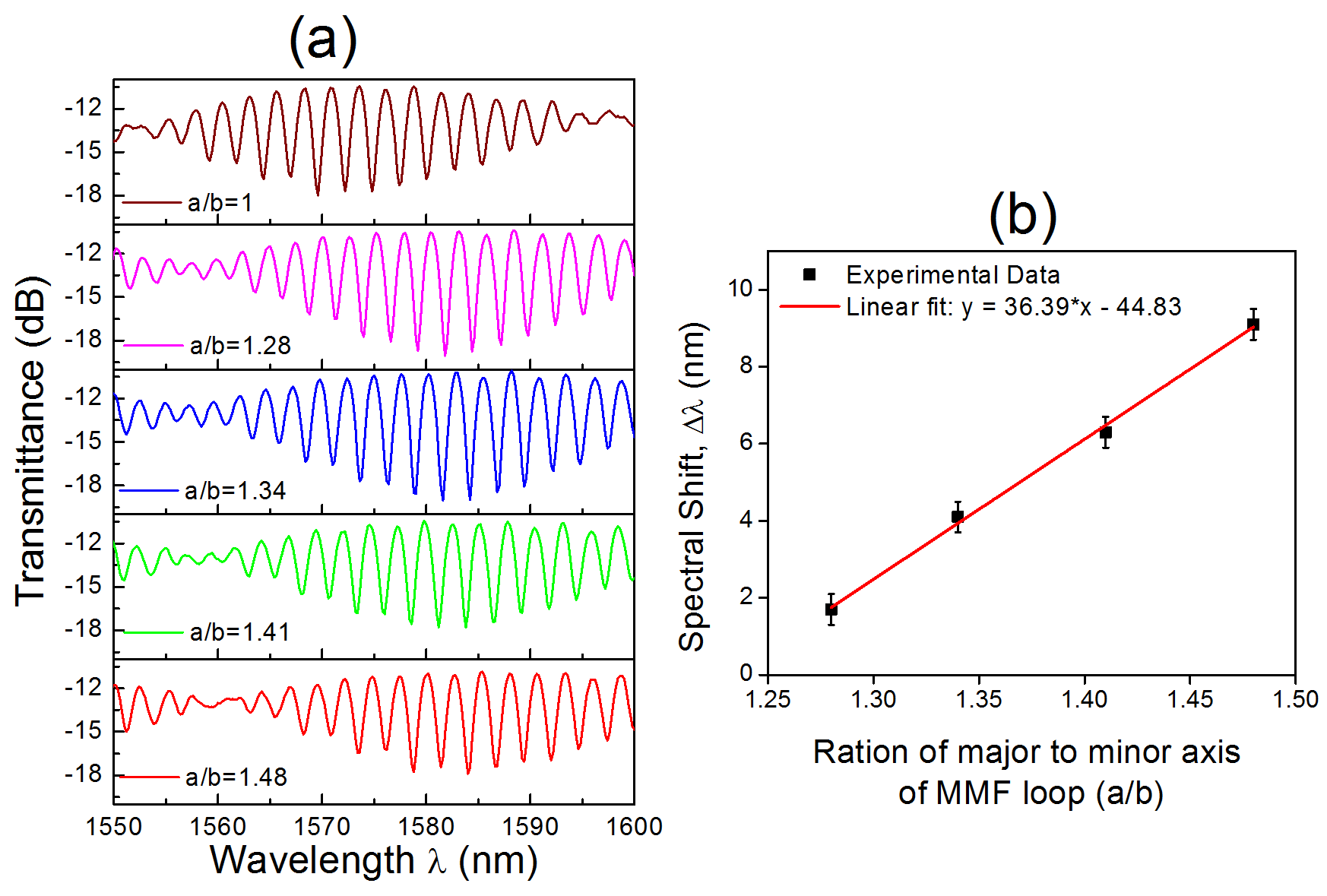}
\end{center}
\caption {(a)Transmission spectra of SMS structure and (b) Spectral shift of transmission spectra corresponding to MMF loop axis ratio, $a/b$, for the wavelength range 1550-1600 nm.}
\label{loopratio}
\end{figure} 

Next, to demonstrate the tuning capability of the filter we recorded the transmission spectrum by changing the MMF fibre loop major to minor axis ratio ($a/b$). The transmission spectra for five different $a/b$ values of 1.00, 1.28, 1.34, 1.41 and 1.48 are shown in Fig. \ref{loopratio}(a). With increasing $a/b$ we observed a blue shift in transmission spectrum as shown in Fig. \ref{loopratio}(a). In contrast to heat based tuning of WDM channel isolation filter as reported by Tiwari et al. \cite{Tiwari}, in the present study tuning of the WDM isolation filter was achieved by the major to minor axis ratio of the MMF loop.
 
In Fig. \ref{loopratio}(b), we have plotted the experimentally observed spectral shift ($\Delta\lambda$) of the transmission minima with the ratio of major to minor axis ($a/b$) of the MMF loop over a wavelength range of 1550 $nm$ – 1600 $nm$. The spectral shift was obtained by taking the spectra corresponding to $a/b$ = 1 as reference. Evidently the spectral shift shows a linear response throughout $a/b$ variation in range from 1.28 to 1.48. A little deviation from linear tuning can be attributed to the errors associated with minimal error in estimation of $a/b$.

\begin{figure}
	\begin{center}
		\includegraphics [width=9 cm]{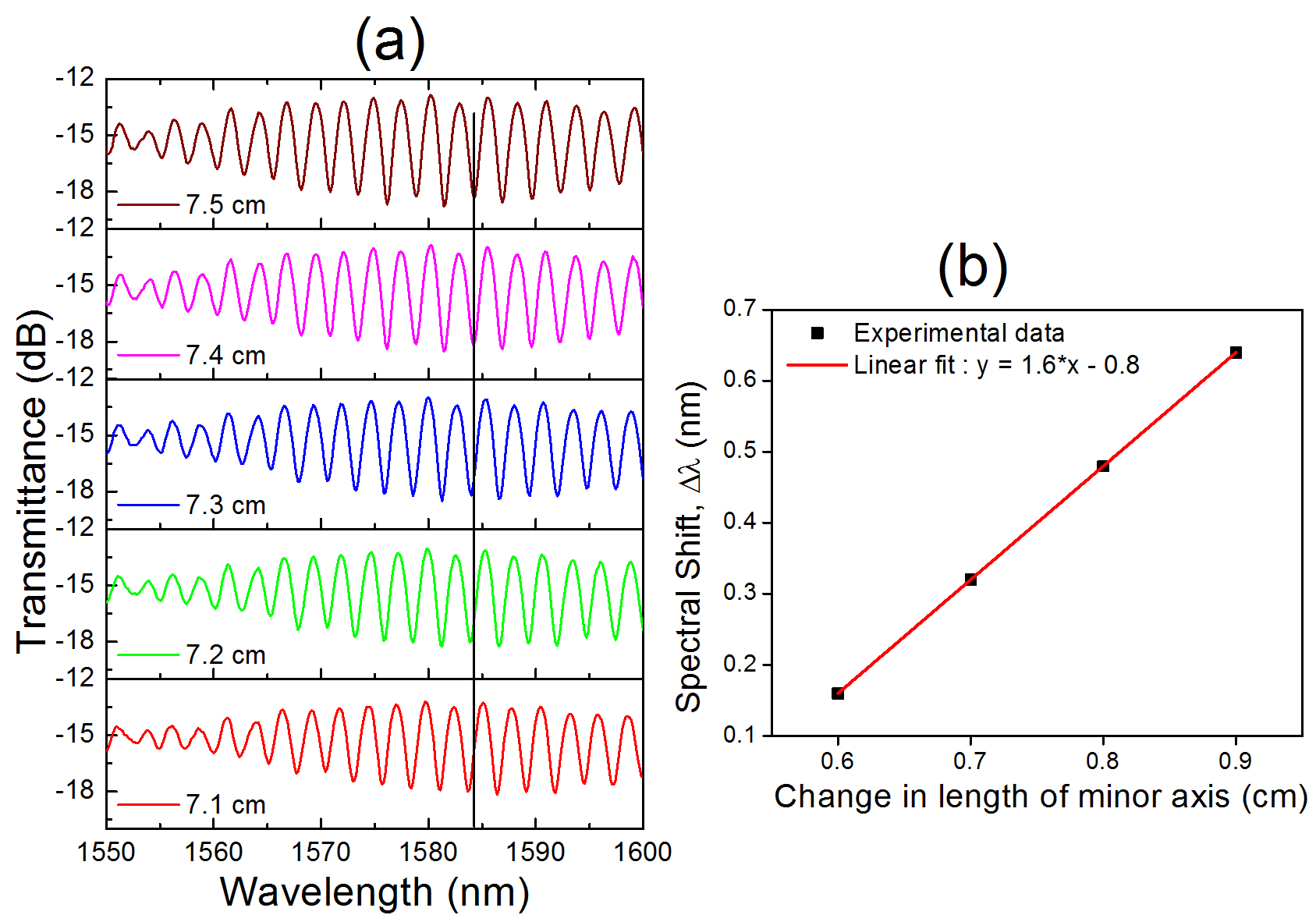}
	\end{center}
	\caption {(a)Transmission spectra of SMS structure and (b) Spectral shift corresponding to change in length of minor axis.}
	\label{verticalpress}
\end{figure}

In another arrangement, the SMS structure was mounted on rigid point and force was applied vertically along the loop axis. The length of loop axis was decreased along the applied force direction and there is a change in the major to minor axis of the MMF. The transmiaaion spectra were recorded for change in axis length from 7.5 $cm$ to 7.1 $cm$ and shown in Fig. \ref{verticalpress}(a). The loop experiences a pressure $mg \Delta h/A$, where, $m$ is mass the loop, $g$ is acceleration due to gravity, $\Delta h$ is the change in length of loop axis in vertical direction and $A$ is the contact area. The change in pressure on the loop is reflected by the spectral shift in the transmission minima, as shown in Fig. \ref{verticalpress}(b).

\begin{figure}
	\begin{center}
		\includegraphics [width=9 cm]{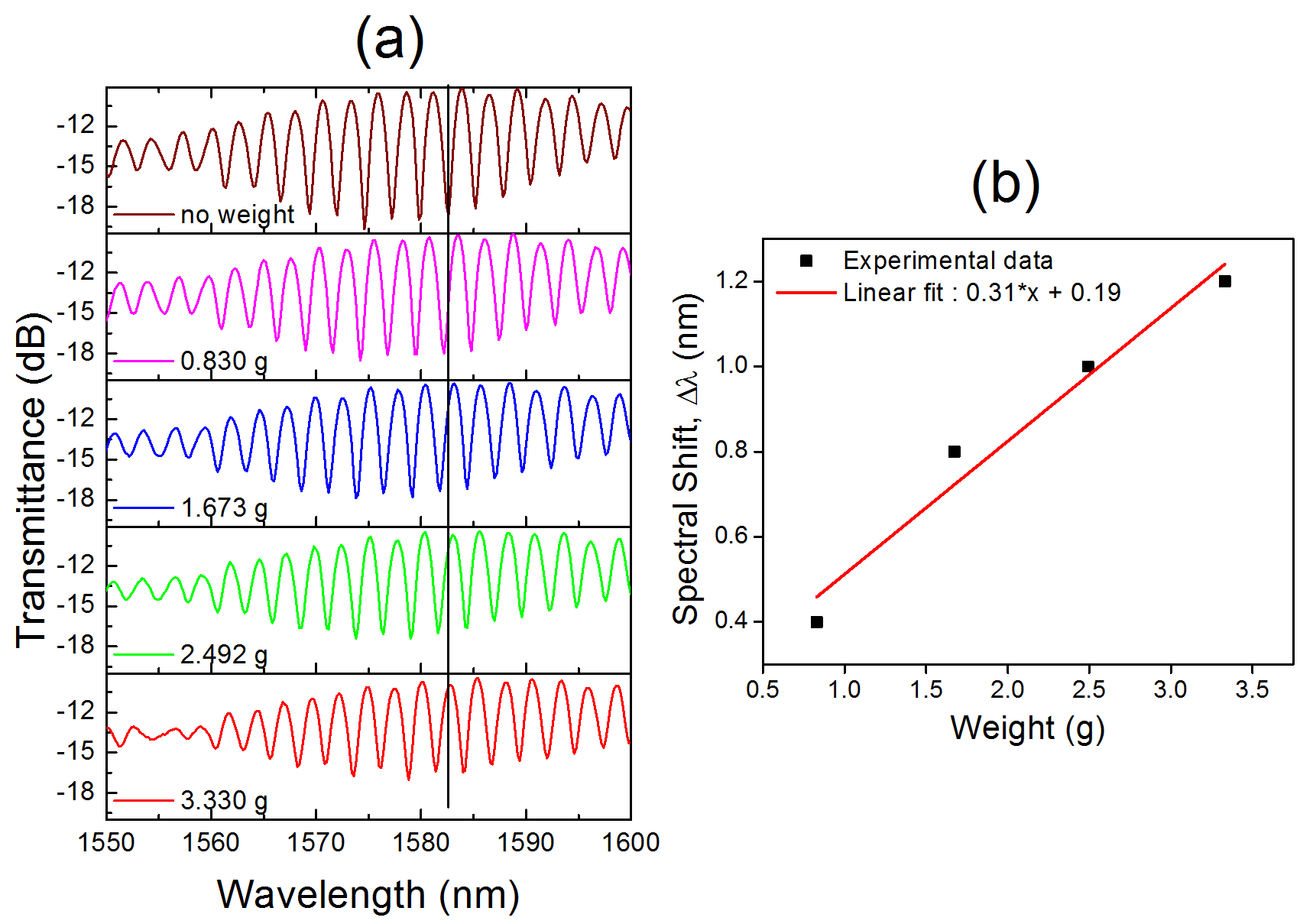}
	\end{center}
	\caption {(a)Transmission spectra of SMS structure and (b) Spectral shift corresponding to various weights.}
	\label{verticalweight}
\end{figure}

Further we carried out an experiment by mounting the loop vertically with a support and hanging some weight to it. Due to weight, one of the loop axis was elongated which leads to the increase in major to minor axis ratio. The transmission spectra were recorded at different weights; 0.830, 0.1673, 0.2492 and 0.3330 $g$ for a wavelength range 1550-1600 $nm$, shown in Fig. \ref{verticalweight}(a). The sensitivity of the SMS structure was studied by the spectra shift of the transmission minima, Fig \ref{verticalweight}(b), and found to be 0.31 $nm/g$ . 

\begin{figure}[h!]
	\begin{center}
		\includegraphics [width=9 cm]{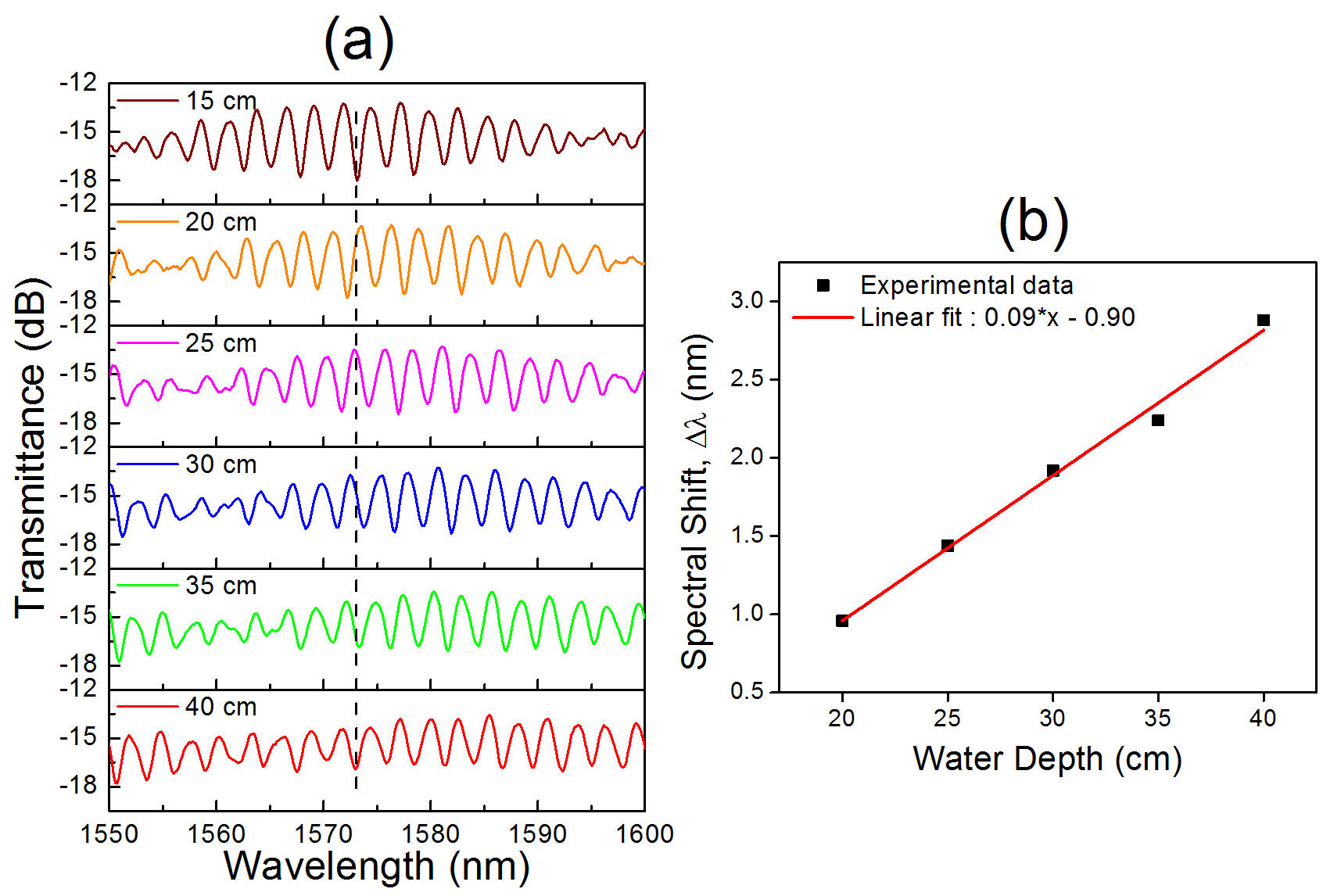}
	\end{center}
	\caption {(a) Transmission spectra of SMS structure and (b) Spectral shift at various water depths.}
	\label{depth} 
\end{figure}

The change in extinction ratio with the major to minor axis ratio of proposed SMS structure can be used for water depth sensor. To demonstrate this, we carried out an experiment by placing the SMS structure vertically in a water container. The SMS structure was dipped in and transmission spectra was measured at water depth ranging 15-40 $cm$ in the interval of 5 $cm$. The upper and lower portion of the MMF loop experiences the different pressure and resulting into the change in major to minor axis ratio of the loop. This effect was reflected in transmission spectra as a spectral shift with the water depth, plotted in Fig. \ref{depth}(a). The spectral shift of the transmission minima with the water depth is shown in Fig. \ref{depth}(b) and the sensitivity was found to be 0.09 $nm/cm$.

\begin{figure}[h!]
	\begin{center}
		\includegraphics [width=9 cm]{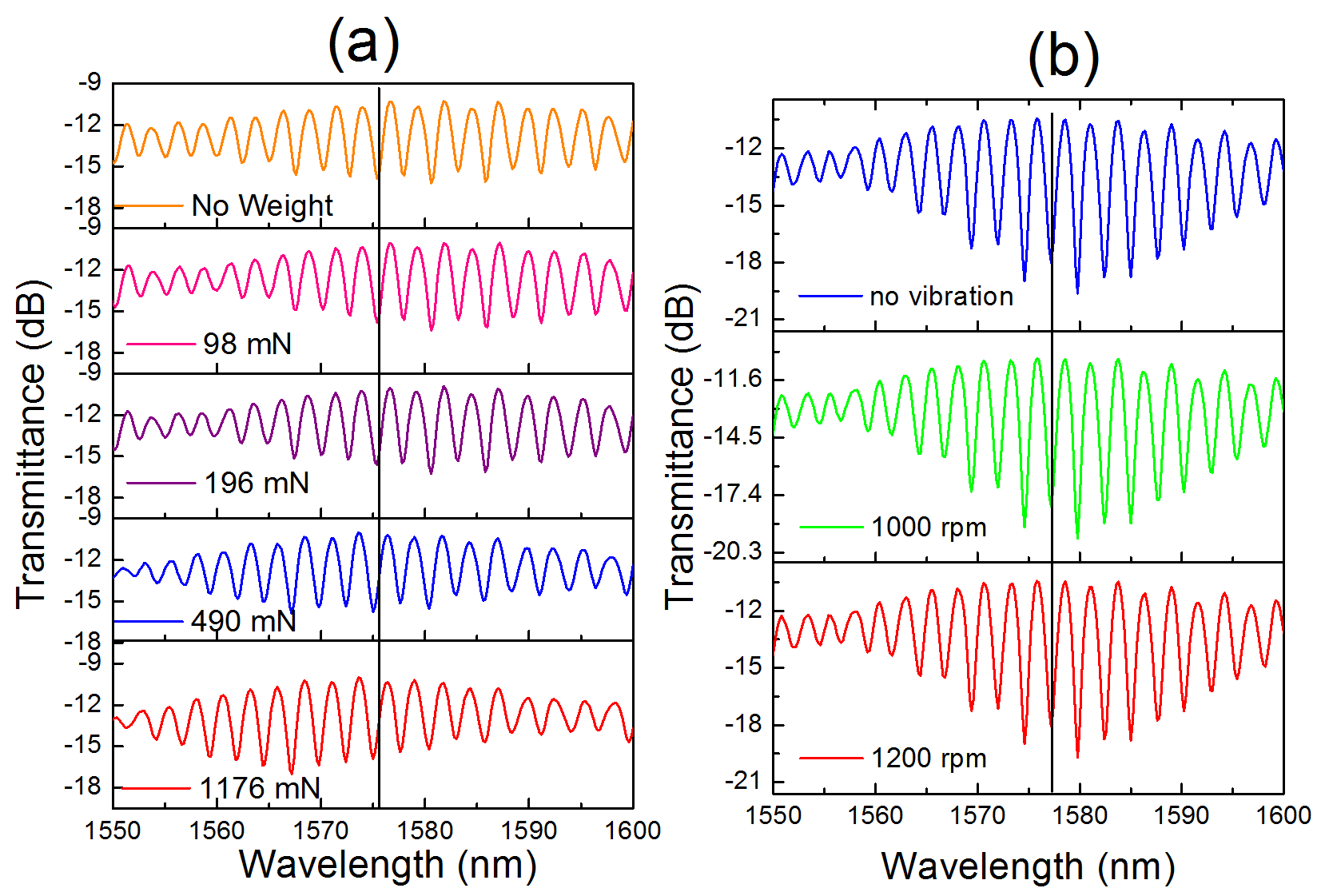}
	\end{center}
	\caption {Transmission spectra of SMS structure (a) when weight are placed on it and (b) when it was kept on vibration plate.}
	\label{weight} 
\end{figure}

We performed another experiment in order to study the response of the proposed SMS structure with pressure when it was kept horizontally on a flat surface. The pressure was applied by placing weight onto structure. The transmission spectra were recorded for the various weights; 98, 196, 490 and 1176 $mN$, shown in Fig. \ref{weight}(a). We observed the spectral shift of the transmission minima with weight but it is too small for using the SMS structure as a pressure sensor in this range. The vibration response of the SMS structure was also studied. For this, the structure was kept on a vibration plate and the transmission spectra were recorded. It can be observed from Fig. \ref{weight}(b) that there is hardly any spectral shift of the transmission minima upto the 1200 rpm.  

\section{Theoretical Analysis}

To understand the experimental results more clearly we also carried out theoretical study with appropriate theoretical model. 

Let us represent the fundamental modal field of the SMF as $\psi_S(r)$ and the modal field for the MMF fibre as $\psi_M(r)$ (= $\sum_{i} a_i \psi_i$), where $a_i$ and $\psi_i$ are the field amplitudes at the splice points and field of the axially symmetric $i^{th}$ mode of the MMF. The field amplitude ($a_i$) is the overlap integral between the fundamental mode of the SMF and $i^{th}$ mode of the MMF, which basically governs the coupling efficiency of various modes of MMF and SMF \cite{Kogelnik, Meunier} and it is expressed as \cite{Ghatak}:

\begin{equation}
     a_i = 2 \pi \int \psi_S(r) \psi_M^i (r) dr
\end{equation}

 In our analysis the transverse offset between the SMF-MMF at the input and output splice points has been incorporated by replacing $\psi_s(r)$ with $\psi_s(r - r_{in})$ and $\psi_s(r - r_{out})$, respectively. Here, $r_{in}$ and $r_{out}$ are the transverse offsets at the input and output splice, and in accordance with the experimental offset their values are taken as 2.7 $\mu m$ and 1.5 $\mu m$ respectively. The normalised modal fields of the SMF and MMF has been taken as \cite{Ghatak, Marcuse}  

\begin{equation}
     \psi_S(r) = \sqrt{\frac{2}{\pi}} \frac{1}{W_s} e^{-\frac{r^2}{W_s^2}}
\end{equation}
\begin{equation}
    \label{simple_equation}
     \psi_M(r) = \sqrt{\frac{2}{\pi}} \frac{1}{W_m} L_m\bigg(\frac{2 r^2}{W_m^2}\bigg) e^{-\frac{r^2}{W_m^2}}
\end{equation}

The total power ($P_{SM}$) carried by the lead out SMF is expressed as 
\begin{equation}
P_{SM}=|a_0^{in} a_0^{out} + a_1^{in} a_1^{out} e^{i(\beta_0-\beta_1)L} + a_2^{in} a_2^{out} e^{i(\beta_0 -\beta_2)L} + ...|^2
\end{equation}

Here, $a_m^{in/out}$ is the overlap integral (calculated using eqn. 1) of the $LP_{01}$ mode of SMF with $LP_{0m}$ modes of the MMF at the lead in/out splice of the SMS structure.
 In accordance with the experimental refractive index profile of the MMF, in our calculations also we considered a parabolic index distrubition of the MMF. The propagation constants corresponding to different modes in the MMF are given by \cite{Ghatak} 
 
\begin{equation}
  \beta_{lm} = k_0 n_0 \bigg[1 - \frac{2(2m +l -1)\alpha_m}{k_0^2 n_0^2}\bigg]^{\frac{1}{2}}  
    \end{equation}

Here, $k_0$ and $n_0$ are the free space wave number and the refractive index of the core of MMF, respectively. The parameter $\alpha_m = 2/W_m^2$, with $W_{m}$ being the spot size of the $m^{th}$ mode of the MMF. For the symmetric modes $l$ = 0 and $m$ has the values 0, 1, 2, etc. 
In our calculation, the core region of both SMF and MMF are considered to be made of $GeO_2$ doped silica with the dopant concentrations of 3.1 mol\% and 13.5 mol\%, respectively. The wavelength dependent refractive indices of core and cladding regions of both the fibres are calculated using the well-known Sellmeier relation \cite{Adams}. In our analysis, we considered only the excitation of the symmetric modes of the MMF as our calculations show that the fractional modal power of $LP_{1m}$ and other anti-symmetric modes is less than 0.003. For the ease of calculation, the bending induced phase variations among the guided modes of the MMF are incorporated in terms of the axial strain between points ``\textit{A}" and ``\textit{B}" (see Fig. \ref{exptsetup}). Due to strain the refractive index of core/cladding and core radius of the fibre are changed and the corresponding changes can be expressed as: \cite{Tripathi}

\begin{figure}[t!]
	\begin{center}
		\includegraphics[width=8 cm]{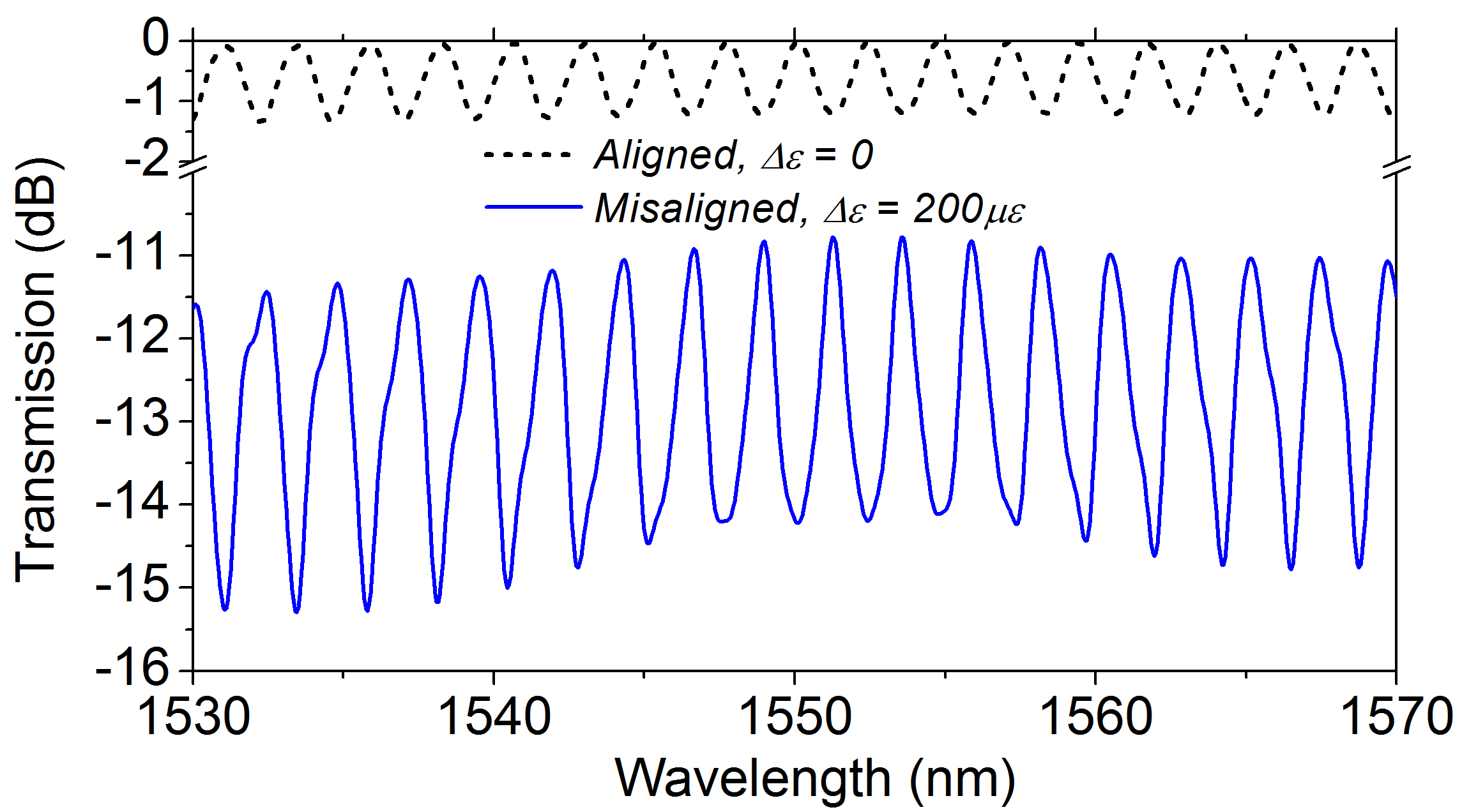}
	\end{center}
	\caption {Simulated transmission spectra of SMS structure for perfectly aligned (black dotted line) and misaligned (blue solid line) cases.}
	\label{misalignsimul}
\end{figure}

\begin{equation}
    \label{simple_equation}
     n_i ~=~ n_{0i} ~+~ \frac{\partial n_{i}}{\partial l} \Delta l~=~n_{0i}- \frac{n_i^3}{2l} [p_{12} - \sigma(p_{11} + p_{12})]\Delta l
\end{equation}

\begin{equation}
    \label{simple_equation}
     a_M ~=~ a_{0M} ~+~\frac{\partial a_{M}}{\partial l} \Delta l
\end{equation}
where, ($\partial a_{M}/ \partial l) \Delta l = (a_M \sigma /l )\Delta l = - a_m \sigma \varepsilon $, $\sigma$ is Poisson ratio, $\varepsilon$ is the axial strain, $n_i$ and $n_{0i}$ is the refractive indices of core/cladding with and without strain, $\Delta l$ is the elongation in the fibre, $p_{11}$ and $p_{12}$ are the strain-optic coefficients of the fused silica. The values of the parameters: $p_{11}$ = 0.12, $p_{12}$ = 0.27, $\sigma$ = 0.17  \cite{Tripathi} and $l$ = 6 $m$ are used in our calculation. The theoretically calculated  transmission spectrum corresponding to the experimental parameters for the axially aligned and misaligned cases (Fig. \ref{misalign}) are shown in Fig. \ref{misalignsimul}. The sinusoidal transmission spectrum suggest that the most of the power is shared between two modes of the MMF. The little deviation from a perfect sinusoidal transmission spectrum can be attributed to the slight excitation of more than two modes of the MMF.

Similar to the experimental results, for the aligned case the extinction ratio of the transmission is quite low (nearly 1.2 $dB$) while in the misaligned case the extinction ratio is enhanced. 
 
\begin{figure}[h!]
\begin{center}
\includegraphics[width=8 cm]{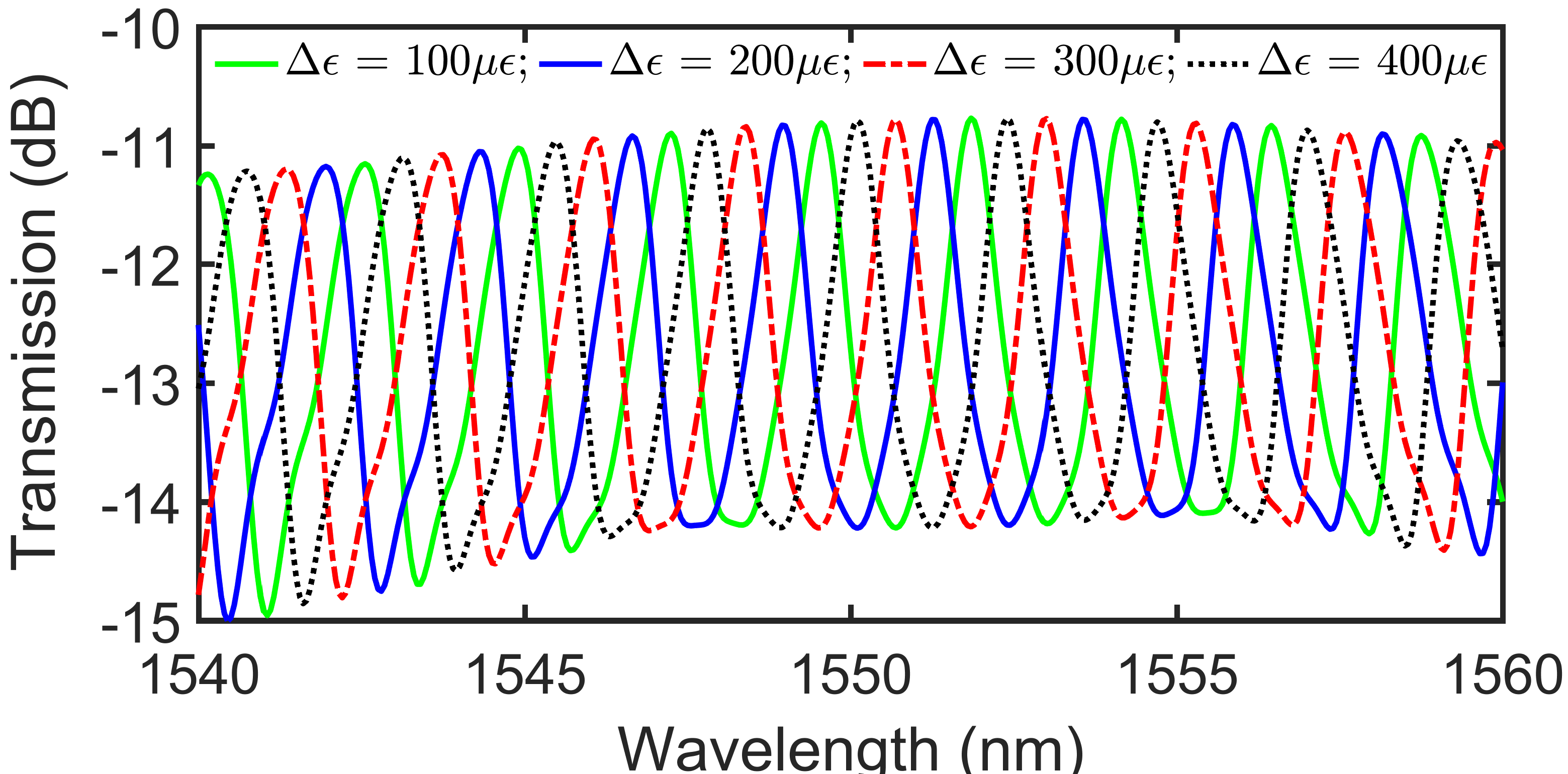}
\end{center}
\caption {\label{Fig8} Simulated transmission spectra of WDM filter for different strains in the MMF loop.}
\end{figure} 

\begin{figure}[h!]
\begin{center}

	\includegraphics[width=7 cm]{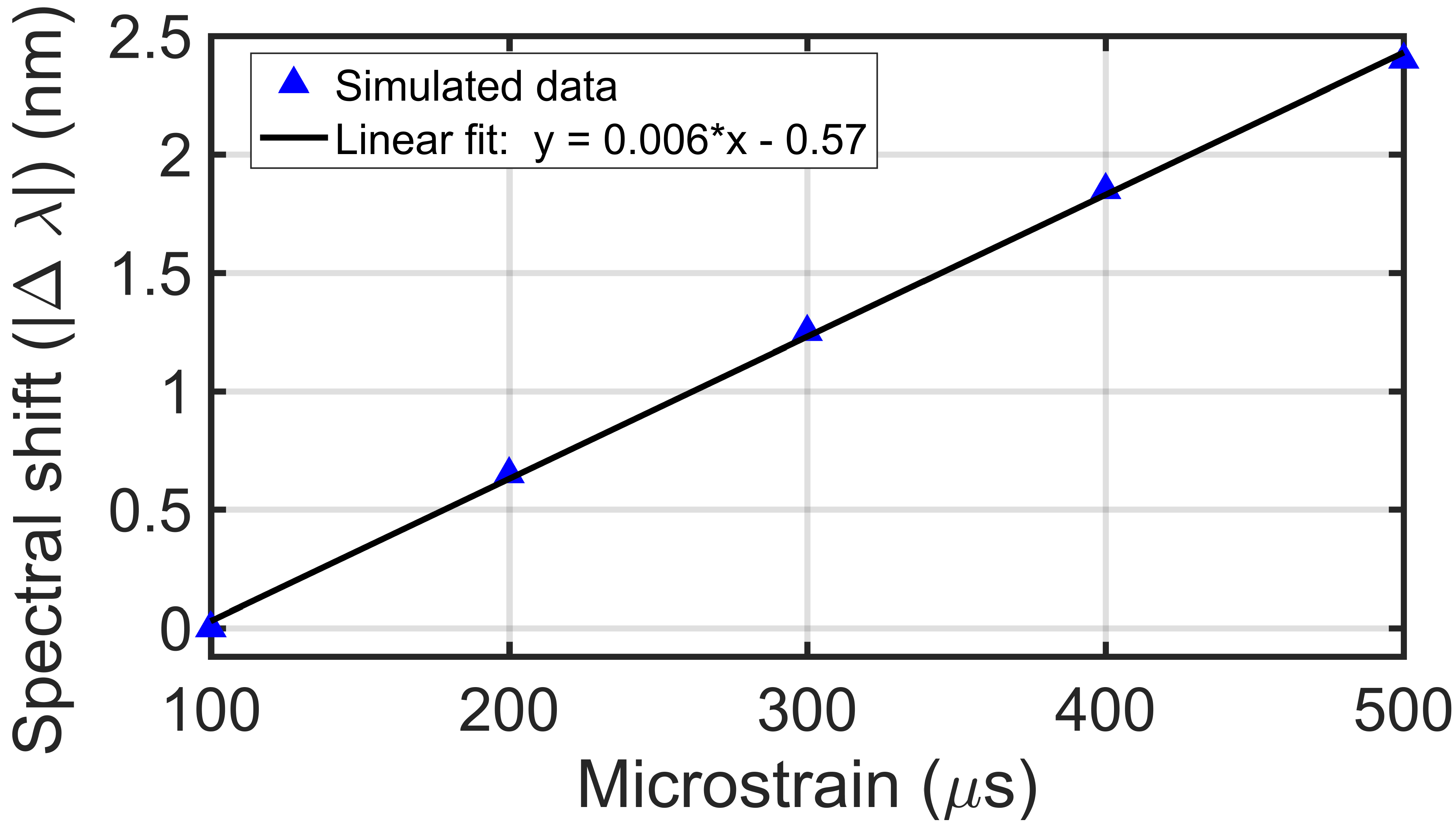}
	
\end{center}
\caption {\label{Fig9} Simulated spectral shift with the strain corresponding to the transmission minima for the wavelength range 1540 nm-1560 nm and its linear fit(black).}
\end{figure} 
To show the tuning behaviour, we increased the microstrain ($\mu s$) from 100 to 400 with the step size of 100. The corresponding transmission spectrum shown in Fig. \ref{Fig8} shows blue shift. In agreement with the experimental results, theoretical spectrum also exhibits a linear blue shift with increasing axial strain, which has been shown in Fig. \ref{Fig9}. The rate of change of the peak/dip position with respect to strain is nearly 0.006 $nm/\mu\epsilon$ (green to black line in Fig. \ref{Fig8}).

\begin{figure}[h!]
\begin{center}
\includegraphics [width=9 cm]{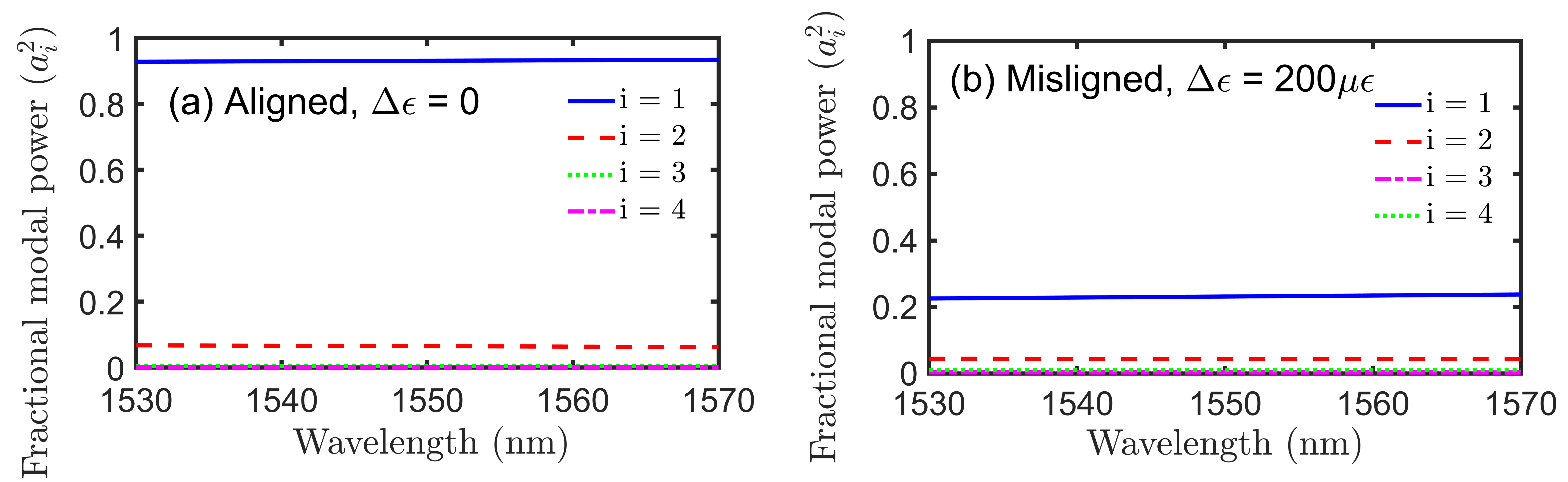}
\end{center}
\caption {Fractional modal power carried out through few modes of the MMF connected between two SMFs for (a) aligned and (b) misaligned cases.}
\label{Fig11}
\end{figure} 

To theoretically demonstrate the underlying mechanism of increased extinction ratio after inserting the misalignments, the normalised fractional modal powers of different modes of the MMF for aligned and misaligned cases are plotted in the Fig. \ref{Fig11}. Here we notice that for the aligned case most of the power is carried by fundamental mode ($a_1^2$) and the ratio of fractional power in fundamental mode to all other higher order modes of the MMF ( $a_2^2, a_3^2, a_4^2 $, etc) is nearly 10 ( Fig. \ref{Fig11}(a)). Hence the modal interference results a low extinction ratio ( dotted curve in Fig. \ref{misalignsimul}). On the other hand for the misaligned case, the fractional power shared in the fundamental mode is reduced, bring it much closer to the fractional modal power carried by $LP_{02}$ and other higher order modes( Fig. \ref{Fig11}(b)). Hence the interfered output of the lead-out SMF exhibits an enhancement in the extinction ratio( solid curve in Fig. \ref{misalignsimul}). 

\section{Conclusion}
In conclusion, we present a new technique for water pressure measurement based on multi-modal interference effect in optical fibers by using single mode-multimode-single mode structure. This method is cost effective than other existing methods and is very easy to fabricate. Theoretical calculations are also carried out to explain the increased extinction ratio. A fair agreement is found between the theory and experimental results.

\end{document}